\documentclass[12pt]{iopart}
\usepackage{epsf,graphicx,amssymb,verbatim,color,cleveref}

\begin{document}
\title[Multilayer coevolution dynamics of the nonlinear voter model]
{Multilayer coevolution dynamics of the nonlinear voter model}
\author{Byungjoon Min$^{1,2}$ and Maxi San Miguel$^{1,*}$}
\address{$^1$IFISC, Instituto de F\'isica Interdisciplinar y Sistemas Complejos (CSIC-UIB),
	Campus Universitat de les Illes Balears, 07122 Palma de Mallorca, Spain}
\address{$^2$Department of Physics, Chungbuk National University, Cheongju, Chungbuk 28644, Korea}
\ead{$^*$maxi@ifisc.uib-csic.es}

\begin{abstract}
We study a coevolving nonlinear voter model on a two-layer network. Coevolution stands 
for coupled dynamics of the state of the nodes and of the topology of the network in 
each layer. The plasticity parameter $p$ measures the relative time scale of the 
evolution of the states of the nodes and the evolution of the network by link rewiring. 
Nonlinearity of the interactions is taken into account through a parameter~$q$ that 
describes the nonlinear effect of local majorities, being $q=1$ the marginal situation 
of the ordinary voter model. Finally the connection between the two layers is measured 
by a degree of multiplexing $\ell$. In terms of these three parameters,~$p$, $q$ and $\ell$ 
we find a rich phase diagram with different phases and transitions.
When the two layers have the same plasticity $p$, the fragmentation transition observed 
in a single layer is shifted to larger values of $p$ plasticity, so that multiplexing 
avoids fragmentation. Different plasticities for the two layers lead to new phases 
that do not exist in a coevolving nonlinear voter model in a single layer, namely 
an asymmetric fragmented phase for $q>1$ and an active shattered phase for $q<1$. 
Coupling layers with different types of nonlinearity, $q_1<1$ and $q_2>1$,
we can find two different transitions by increasing the plasticity parameter, 
a first absorbing transition with no fragmentation and a subsequent fragmentation transition.
\end{abstract}
\maketitle

\section{Introduction}

Real-world networked systems ranging from biological systems~\cite{vidal,white}
and human society~\cite{borge,verbrugge} to transportation~\cite{domenico}
and infrastructure systems~\cite{little,rosato,robustness}
are rarely isolated but often formed by multiple layers of networks.
In order to perform functionality properly, the networked systems
maintain multilayer structures and interactions between different layers of networks.
The concept of multilayer networks~\cite{kivela}
has been proposed along with interconnected networks~\cite{leicht},
interdependent networks~\cite{buldyrev},
and multiplex networks~\cite{lee}, for a more complete modeling of interconnected systems.
Multilayer networks are a framework not only for a better description of complex
systems but also for novel dynamical processes that cannot be captured in a single
layer framework~\cite{kivela,boccaletti,wang,marina2}. Indeed, several studies on multilayer
networks show that interlayer connections account for
significant differences in many phenomena,  including
percolation~\cite{leicht,son,viability}, diffusion~\cite{moreno},
epidemic spreading~\cite{switching,zheng,rescue,czaplicka}, cascade of
failures~\cite{buldyrev,brummitt},
opinion formation~\cite{marina2,marina,amato,artime}, online communities \cite{klimek}, 
game theory~\cite{lugo,szolnoki,wang2} or cultural dynamics \cite{battiston}.

One fundamental feature studied in multilayer networks is coevolution
dynamics~\cite{marina,klimek,perc}, that is the evolution of a network structure in response to
the dynamical processes that change the state of the nodes \cite{zimmerman}.
A coevolving voter model is a representative model of coevolution dynamics on complex networks~\cite{vazquez,holme}.
An ordinary coevolving voter model consists of two different kinds of
processes: copying and rewiring. The ratio of time scales at which these two processes occur 
is measured by a parameter $p$ called plasticity of the network. For the copying process, 
with a certain probability $p$ a node changes its state by copying the state of one of its neighbors
randomly chosen, following the original imitation mechanism implemented by the voter model.
For the rewiring process, with the complementary probability $1-p$ a node
rewires its connection with a neighbor having a different state, to another node
having the same state. The ordinary coevolving voter model exhibits
an absorbing phase transition between an dynamically active coexistence phase and an absorbing
phase in the thermodynamic limit~\cite{vazquez}. The finite size manifestation of this 
transition is a network fragmentation transition.
Coevolution dynamics of the voter model on multilayer networks gives a more complete modeling 
of real world situations: for instance, the individuals' opinion and social networks may evolve through
multiple different types of social relationship, such as family, friends, and colleagues, 
or communication, friendship and trade networks. An important parameter in this multilayer description is
the degree of multiplexity measured by the density of interlayer links, i.e.,
density of links between nodes in different layers. It has been found that a coevolving 
voter model in a multilayer network exhibits a shattered fragmentation with a phase 
showing many disjointed small components \cite{marina}. This phase does not exist 
in a coevolving voter model in a single network layer. In addition, it has also 
been shown that the voter model on multilayer networks
cannot be reduced to a single layer description~\cite{marina2}.
Therefore, the structure of multilayer networks significantly affects
the dynamical consequences of coevolution in the voter model~\cite{marina,klimek}.

More recently, collective or group interactions beyond the dyadic interactions of the 
voter model have been considered within a coevolution dynamics context~\cite{cnvm,triadic}.
Specifically, a coevolving nonlinear voter model (CNVM) has been studied
in order to incorporate collective interactions and coevolution dynamics
at the same time~\cite{cnvm}. The nonlinearity in the CNVM takes into account that the state of
an agent is affected by the state of all of their neighbors as a whole, and not by
a pairwise interaction~\cite{castellano,nyczka,mf,peralta}. The nonlinear interaction gives rise to diverse
phases, with different mechanisms for fragmentation transitions.
Such form of nonlinearity was also studied in social impact theory~\cite{nowak}, in 
language evolution problems~\cite{nettle}, or in language competition dynamics under 
the name of volatility~\cite{abrams,jstat,castello}.
However, the effect of the nonlinearity in a coevolving voter model
has been examined only on a single layer network as the simplest example.

In view of the nontrivial modifications found for a coevolving voter model when 
considering a multilayer framework, we address in this paper the study of a 
coevolving nonlinear voter model on a multilayer network.
The outline of the paper is as follows. In section 2 we specify our dynamical model. 
Section 3 describes our results for the case in which the two layers have the same 
plasticity $p$. In this case we find that the fragmentation
transition found in a single layer \cite{cnvm} continues to exist, but with a 
delayed threshold of $p$. Our numerical results are qualitatively described 
by a mean-field approach. Section 4 considers the case in which the two 
layers have different plasticities, and we find a rich variety of phases 
such as a dynamically active shattered phase, an asymmetric fragmented phase, and
a coexistence phase. In section 5 we analyze the case of layers with different 
nonlinearity which also results in other non-trivial phase transitions that are 
not observed in a single layer network. For instance, 
two subsequent transitions can occur among coexistence, consensus,
and absorbing fragmented phases.

\section{Model}

\begin{figure}[t]
\includegraphics[width=\linewidth]{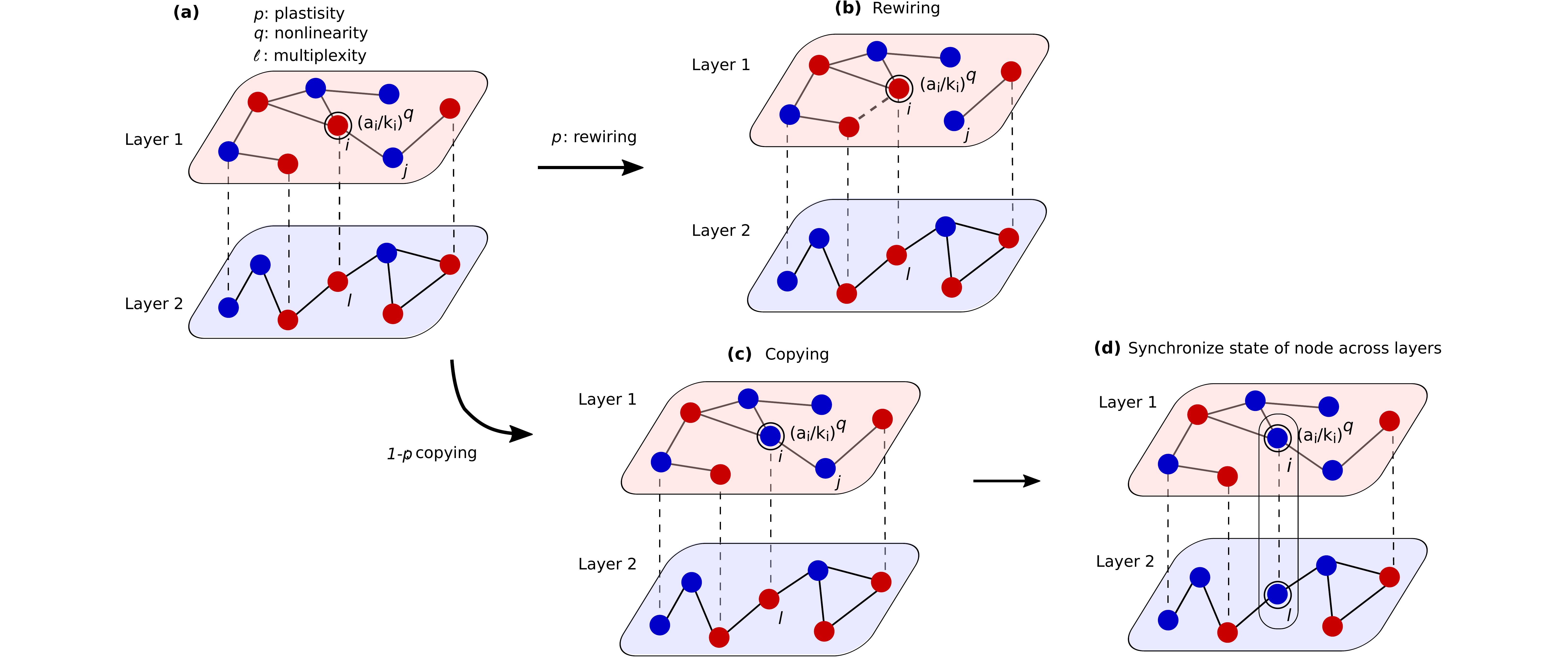}
\caption{Schematic illustration of update rules in a nonlinear coevolving voter
model on multilayer networks. At each step, a layer and node $i$ in the layer are
chosen randomly. Then, with a probability $(\frac{a_i}{k_i})^q$ one of the active 
links is chosen. And, we rewire an active link with probability $p$ and copy the 
state of the chosen neighbor with probability $1-p$. If node $i$ is connected to 
a node in the other layer, the state of the node in the other layer is
synchronized to its state with the state of $i$.
}
\label{fig:model}
\end{figure}

Our model considers multilayer networks composed of two different layers
in which each layer is initially independently constructed as a degree regular network
with the same number of nodes $N$ and with the same number $\langle k\rangle=4$ 
of random intralayer links for each node. Inter-layer links connect two nodes
that belong to the two different layers~(Fig.~\ref{fig:model}).
We define the degree of multiplexity $\ell$ ~\cite{marina} as the density of 
inter-layer links so that $\ell N$ is the total number of links connecting nodes in different layers.
Initially each node $i$ is in one of two states, $s_i=+1$ (up)
or $-1$ (down), with the same probability $1/2$, and it has the same state in both layers.
At a given configuration, links between two nodes in the same layer (intra-links) can be 
classified as active or inert, depending on the state of the pair of connected nodes. 
Active (inert) links stand for the links connecting two nodes in different (same) states.

The dynamical model is as follows: at each step, we randomly choose a layer 
and a node $i$ in the chosen layer. We measure the fraction of active links 
of node $i$ with respect to its degree $k_i$, $\left( \frac{a_i}{k_i}\right)$ 
where $a_i$ is the number of active links of node $i$. Nonlinear interactions 
are implemented through a probability $\left( \frac{a_i}{k_i}\right)^q$, 
where $q$ is the nonlinearity parameter measuring the nonlinear effect of 
local majorities. With this probability, the node $i$ takes an action of either 
copying or rewiring. Then, we choose one of its neighbors~$j$, having
a state different than the one of node $i$ (or equivalently we choose one of 
the active links of node $i$). Note that with the complementary probability 
$1-\left( \frac{a_i}{k_i}\right)^q$, nothing happens and another node is randomly selected.
Next, rewiring occurs with probability~$p$: the chosen active link is removed, 
and rewired to a new node having the same state as the state of node $i$. 
And, with probability $1-p$ node $i$ changes its state by copying the state 
of node $j$. Subsequently, a node connected to $i$ in the other layer
by an inter-link also changes its state adopting the same state than node $i$. 
This synchronization process leads to the same state for connected nodes across 
different layers. While the number of nodes and the density of links are constant, 
the network structure and the configuration of the states of the state vary in time. 
These update processes proceed until the system reaches a steady state.

In our model, we have three main parameters: plasticity or rewiring rate $p$,
nonlinearity $q$, and the degree of multiplexity $\ell$.
First, the plasticity measures how often the process of rewiring occurs
as compared to the process of copying. When $p=0$, a network is static,
so that the model becomes the voter model on multilayer networks \cite{marina2}.
On the other hand, when $p$ is non-zero, both the structure of the network
and the state of the nodes in the network change in a coevolution dynamics. 
In the other extreme $p=1$, there is no copying process and the network 
eventually becomes fragmented due to the rewiring processes. Second, 
the nonlinearity parameter~$q$ measures the effect of local group interactions.  
Nonlinearity is mathematically implemented as 
$\left( \frac{a_i}{k_i}\right)^q$~\cite{cnvm,castellano,nyczka,mf,peralta}. 
When $q=1$, our model becomes the 
ordinary coevolving linear voter model \cite{vazquez}. For $q>1$, nodes with 
more active links have a higher probability, as compared to the ordinary linear 
voter model, to take an action than other nodes. When $q<1$, nodes with less 
active links are more likely to take action than in the linear voter model.
Finally, the degree of multiplexity $\ell$ stands for the density of interlayer links. 
$\ell=1$ corresponds to one-to-one connections among the nodes in the two layers, while
$\ell=0$ corresponds to the case of no interconnections, meaning that the two layers 
are isolated. When $0<\ell<1$, the networks on the two layers are interconnected but 
have sparse interconnections than one-to-one connections.

\section{Symmetric plasticity in multilayer networks}

\begin{figure}
\includegraphics[width=\linewidth]{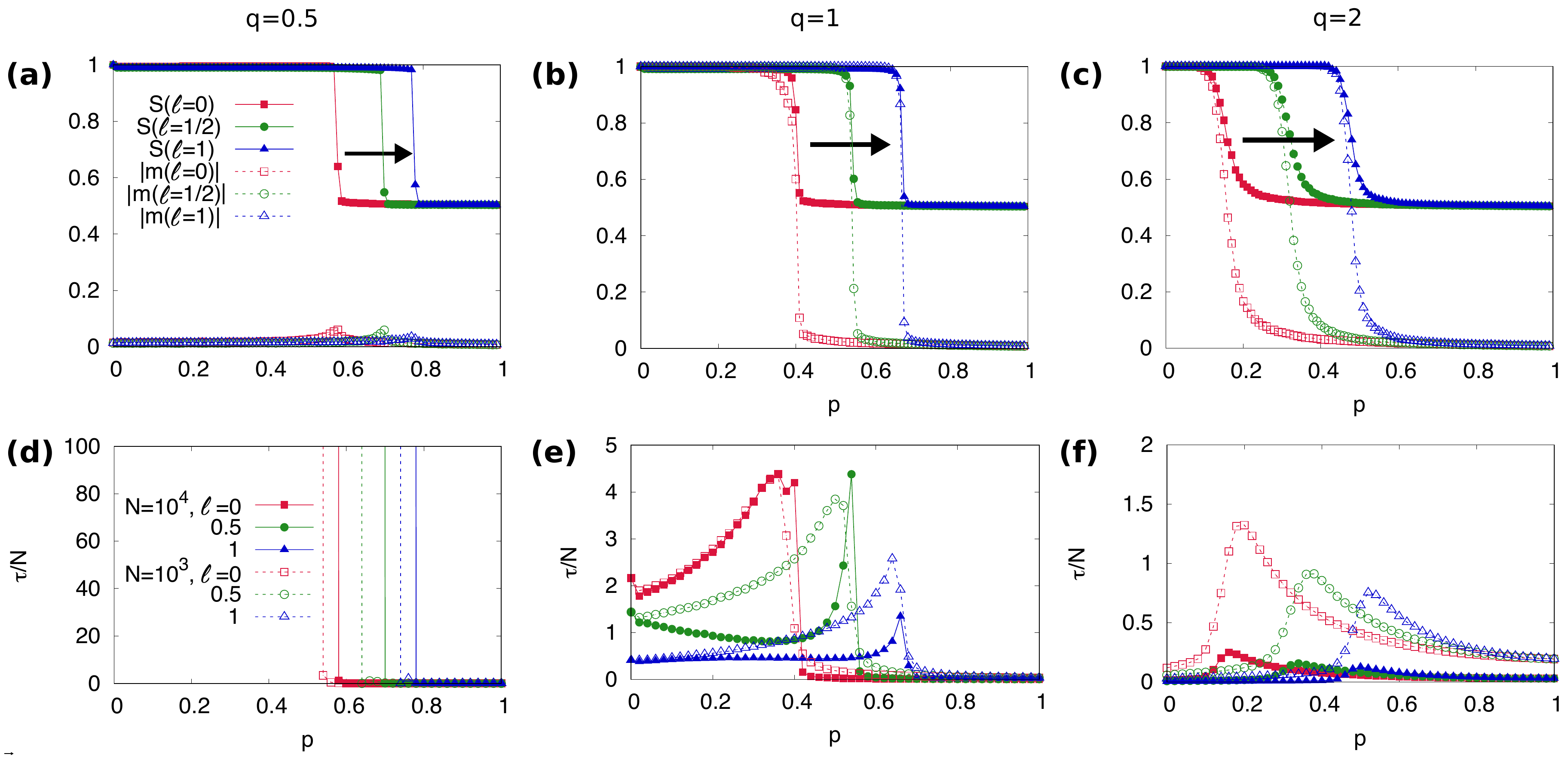}
\caption{
The size $S$ of the largest component and the absolute value $|m|$ of magnetization
of (a) $q=0.5$, (b) $1$, and (c) $2$ with symmetric plasticity $p=p_1=p_2$
for different $\ell=0,0.5,1$ on multilayer degree regular networks with 
$\langle k \rangle =4$ and $N=10^4$ for each layer, averaged over $10^4$ runs.
The characteristics time $\tau$ to reach the final state of 
(d) $q=0.5$, (e) $1$, and (f) $2$ is also shown for $N=10^3$ and $10^4$.
}
\label{fig:sym}
\end{figure}

We indicate as $p_1$ and $p_2$ the plasticities of each layer. The simplest case
of the coevolving model on multilayer networks is the case of symmetric
plasticity, so that $p=p_1=p_2$. To describe the properties of the steady state, 
we use the size $S$ of the largest network component for measuring the global 
connectivity and the absolute value~$|m|$ of magnetization $m=\sum_i s_i$
for each layer. For a single layer, the nonlinearity~$q$ significantly changes 
the coevolution dynamics \cite{cnvm}: when $q<1$, a fragmentation transition 
between a dynamically active coexistence phase in a single component network 
and a fragmented phase occurs for a critical plasticity $p_c$,
while when $q>1$ a distinct type of a fragmentation transition occurs between
an absorbing consensus phase and a fragmented phase.
When $q=1$, which is the case of the ordinary linear voter model, an absorbing phase 
transition between a dynamically active coexistence phase and an absorbing phase 
in a fragmented network is recovered \cite{vazquez,cnvm}.

The same phases and transitions described for the single layer case continue to 
exist in multi-layer networks but the critical plasticity $p_c$ is delayed as 
the degree of multiplexity $\ell$ increases. For different nonlinearity 
parameters $q=0.5$, $1$, and $2$, we determine $S$, $|m|$,
and the characteristic time $\tau$ to reach a final state (Fig.~\ref{fig:sym}).
Note that $S$, $|m|$, and $\tau$ for both layers are
statistically the same due to the symmetric case analyzed here.
When $q=0.5$, we find an absorbing phase transition
between a coexistence phase and a fragmented phase. The coexistence phase, which 
is dynamically active, is well characterized by the divergence of $\tau$ in 
the thermodynamic limit $N \rightarrow \infty$. The fragmented phase corresponds 
to $S=1/2$ and $m=0$, implying two disjoint clusters, each of them in a consensus 
state but but with opposite consensus states. When $q=2$, we find a different 
transition at the critical plasticity $p_c$ between two absorbing phases, 
a consensus and a fragmented phase. The consensus phase is characterized 
by $S=1$ and $|m|=1$, implying a single network component with an ordered state.
For the linear case $q=1$ \cite{marina}, we also observe a delay of the 
fragmentation transition when increasing the degree of multiplexity.

\begin{figure}
\includegraphics[width=\linewidth]{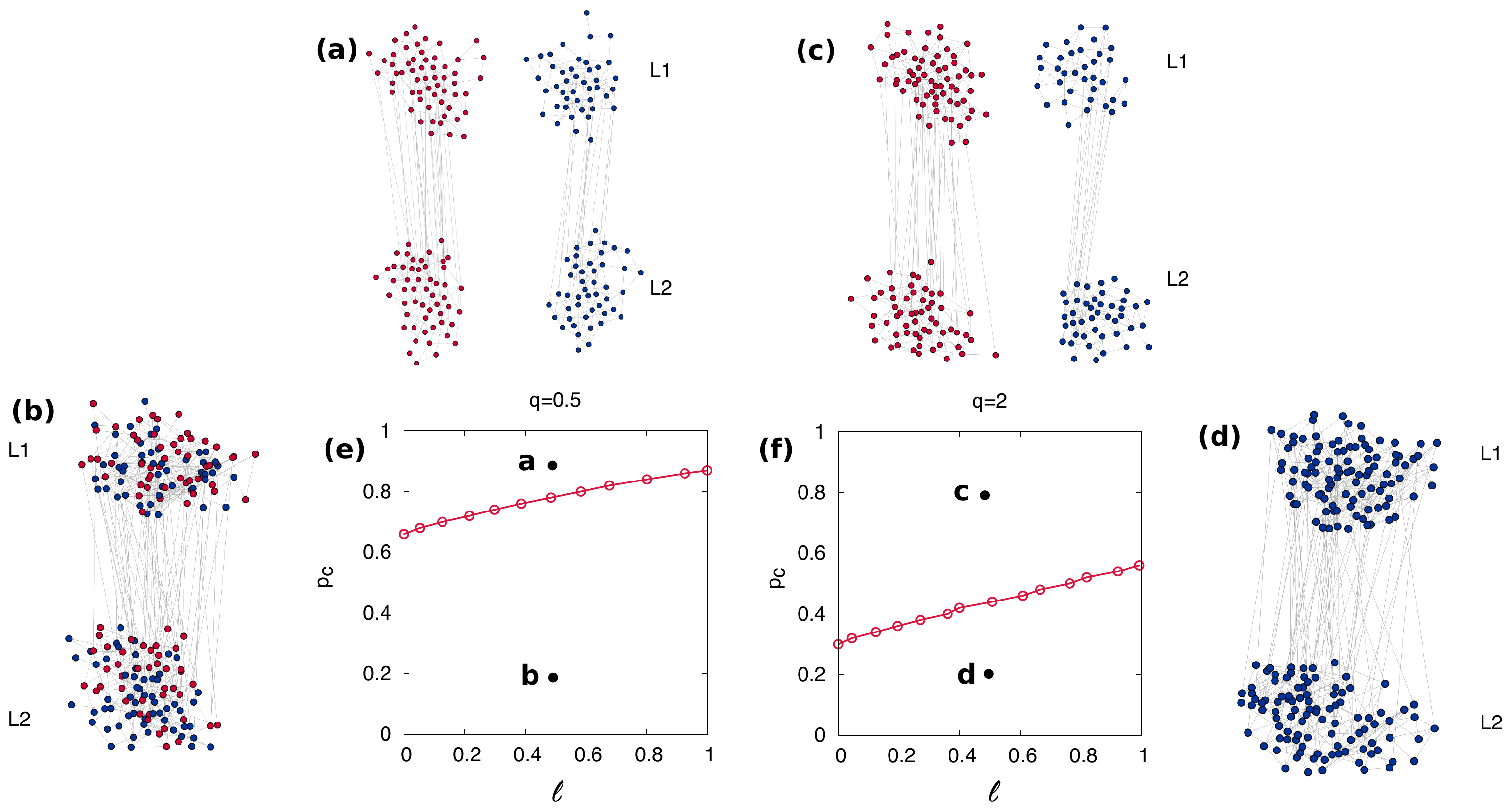}
\caption{
Phase diagram for the symmetric plasticity case with (e) $q=0.5$ and (f) $q=2$
together with network examples of (a) fragmented $(p,q,\ell)=(0.9,0.5,0.5)$, 
(b) coexistence ($0.2,0.5,0.5$), (c) fragmented $(0.8,2,0.5)$, and (d) consensus $(0.2,2,0.5)$ phases.
The phase diagram is obtained numerically with initial multilayer degree regular
networks with $\langle k \rangle =4$ and $N=10^3$. The lines in the diagram 
are just to guide the eyes.
}
\label{fig:phase_p}
\end{figure}

We further examine the effect of the multiplexity in terms of the dependence $p_c$
on $\ell$ for $q=0.5$ and $2$ (Fig.~\ref{fig:phase_p}).
For both $q=0.5$ and $2$, we find a delayed onset~(larger $p_c$) of the fragmented phase
with increasing multiplexity $\ell$. The shift of $p_c$
means that the inter-layer connections in a multilayer structure prolong
the global connectivity in the coevolution dynamics:
multiplexity provides a source of disorder that prevents reaching consensus 
due to the synchronization process. Therefore increasing the degree of multiplexity $\ell$ 
leads to the shift of $p_c$. This the same mechanism that for the linear voter model and 
therefore both for linear and nonlinear interactions, multiplexity prevents fragmentation.
However, the role of nonlinearity is shown in the final state: a dynamically active coexistence
phase in $q=0.5$ and a consensus phase in $q=2$.

In order to obtain an analytic insight for the shift of $p_c$, we introduce mean-field
equations \cite{marina,vazquez,cnvm,mf} for the nonlinear voter model on multilayer 
networks. These equations are valid in the thermodynamic limit $N\rightarrow\infty$.
We define the average degree of each layer as $\langle k_1 \rangle$ and $\langle k_2 \rangle$.
The density of active links $\rho_i$ in each layer $i \in \{ 1,2 \}$
can be described by the following equations \cite{cnvm,mf},
\begin{eqnarray}
\frac{d\rho_i}{dt} = - \rho_i^q p_i + \rho_i^q (1-p_i) \left[ \langle k_i \rangle - 2q -2( \langle k_i \rangle -q)\rho_i \right]
	+ \ell \rho_j^q (1-2 \rho_i) \langle k_i \rangle.
\end{eqnarray}
Note that these coupled equations reduce to previous results in the
appropriate limit of linear interactions $q=1$ \cite{marina} or decoupled 
layers $\ell=0$ \cite{cnvm}. Assuming that the two layers have the same 
mean degree $(\langle k \rangle= \langle k_1 \rangle = \langle k_2 \rangle)$,
and for  symmetric coupling~($p=p_1=p_2$),
\begin{eqnarray}
\frac{d\rho}{dt} = - \rho^q p + \rho^q (1-p) \left[ \langle k \rangle - 2q -2(\langle k \rangle  -q)\rho \right]
	+ \ell \rho^q (1-2 \rho) \langle k \rangle,
\label{eq:rho}
\end{eqnarray}
where $\rho=\rho_1=\rho_2$ due to the symmetry.
For the steady state, the trivial solution $\rho=0$ corresponds
to a fragmentation state where the dynamics is frozen and
the non-zero solution corresponds to a dynamically active state.
The non-zero solution $\rho^*$ of Eq.~\ref{eq:rho} gives
\begin{eqnarray}
\rho^*=\frac{\langle k \rangle(1+\ell-p)-2q+p(2q-1)}{2\langle k \rangle(1+\ell-p)+(1-p)q}.
\end{eqnarray}
The transition point to the fragmentation phase with $\rho=0$ is
\begin{eqnarray}
p_c=\frac{\langle k \rangle+ \ell \langle k \rangle-2q}{1+\langle k \rangle-2q},
\end{eqnarray}
Thus, this mean-field approximation accounts for the linear growth
of $p_c$ with respect to $\ell$ obtained numerically. The analytical approach
predicts successfully the shift of $p_c$ but it gives quantitatively inaccurate values of $p_c$
as it is also the case for the linear voter model \cite{marina}.

\section{Asymmetric plasticity in multilayer networks}

\begin{figure}
\includegraphics[width=\linewidth]{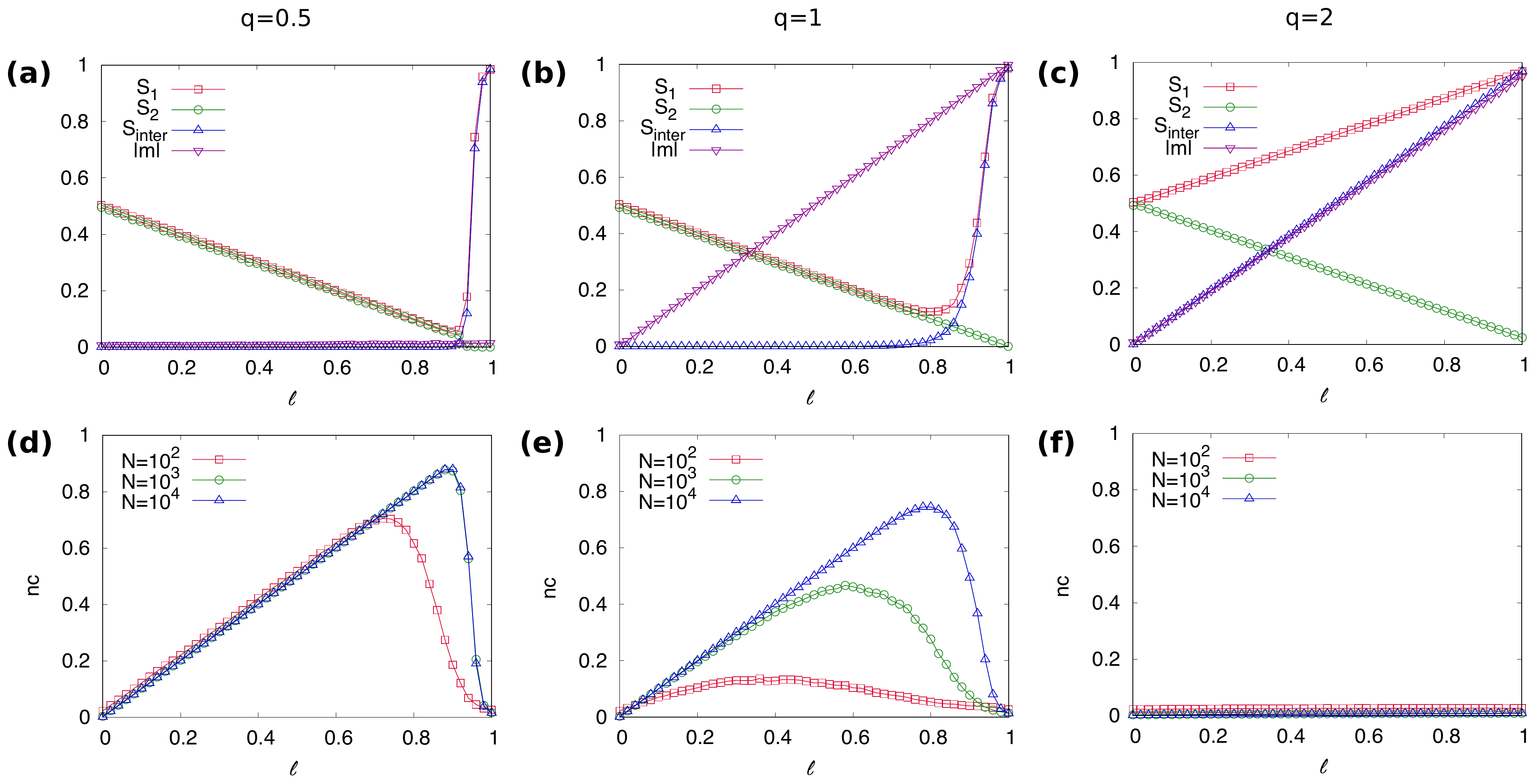}
\caption{
For a bilayer with fully asymmetric plasticity $p_1=1$ and $p_2=0$ the following quantities 
for the dynamic layer $p_1=1$ are shown: size $S_1$ of the largest component, size $S_2$ of 
the second largest component, the fraction $S_{inter}$ of nodes having interconnections 
in the largest network component and the absolute value $|m|$ of magnetization for
(a) $q=0.5$, (b) $1$, and (c) $2$. Numerical simulations on multilayer initially 
degree regular networks with $\langle k \rangle =4$ and $N=10^4$ for each layer.
The relative number of network components $nc$ of the dynamic layer for 
(d) $q=0.5$, (d) $1$, and (f) $2$ is also reported for different system sizes.
}
\label{fig:fully}
\end{figure}

Far from the symmetric case, an extreme coupling scenario is that of fully asymmetric
plasticity, meaning that one layer only rewires ($p_1=1$, the dynamic layer)
and the other layer only changes the state of the nodes ($p_2=0$, the voter layer).
The dynamic layer is affected by the voter layer due to the synchronization step,
but the voter layer is independent of the dynamic layer. Hence as $t \rightarrow \infty$,
the voter layer will either remain in an active coexistence phase ($q<1$), except 
for finite size effects, or will reach a consensus phase ($q>1$), as the result of 
the single layer dynamics \cite{cnvm}. However, the dynamic layer can show a variety 
of asymptotic states depending on the nonlinearity $q$ and the degree of multiplexity $\ell$.
In order to describe these possible states, we determine in the dynamic layer, and for 
$q=0.5,1,2$, the size of the largest network component $S_1$, the size of the second 
largest network component $S_2$, the absolute value $|m|$ of magnetization, and the relative 
number of components $nc$ to the network size $N$, as shown in Fig.~\ref{fig:fully}.
In addition, we also determine the fraction $S_{inter}$ of nodes in the dynamic layer
that belong to the largest network component $S_1$ and at the same time are connected
to the voter layer. Since in our model only a fraction of nodes ($\ell N$) have interlayer links,
$S_{inter}$ refers to the fraction of nodes of the largest network component with interlayer links.
Note that $S_{inter} \le S_1$ and $S_{inter} \le \ell N$.

When $q=1$ (the linear voter case)\cite{marina}, we find a shattered phase in the dynamic layer
for a broad range of values of $\ell$, showing two large components in opposite states and
many isolated nodes [Fig.~\ref{fig:fully}(b,e)].
This shattered phase appears because nodes in the voter layer drive
the separation of nodes in the dynamic layer by the synchronization step of the dynamics.

When $q=0.5$, a representative example of $q<1$, the structure of the dynamical layer
evolves to a shattered phase where two significant network components are in opposite states
and many small clusters exist for a wide range of $\ell$ similarly to the linear voter case ($q=1$).
The relative number of components $nc$ clearly identifies the existence of many isolated
nodes in the dynamic layer [Fig.~\ref{fig:fully}(d)]. As $\ell$ increases from zero, 
$nc$ increases linearly and $S_1$ and $S_2$ decreases linearly as well. $S_{inter}$ is nearly
zero for all $\ell$, indicating that the nodes having interlayer links are those which are isolated 
in their layer. However, the absolute value of the magnetization  $|m|$ in the dynamical layer 
remains zero at variance with what happens for $q=1$.  This neutral magnetization is caused by
a dynamically active coexistence phase in the voter layer which perpetually drives
the magnetization to zero. In this sense we name this phase as an active shattered phase, 
while in the shattered phase for $q=1$ all isolated nodes are in the same state, 
which is the state in which the voter layer has reached a consensus.
For large $\ell$, $S_1$ increases and finally all nodes belong to one connected network
component which is in a coexistence phase ($|m|=0$).
In between these two phases, there is a critical value of the degree of multiplexity $\ell_c$
identifying a transition between an active shattered phase and a coexistence phase.

\begin{figure}
\includegraphics[width=\linewidth]{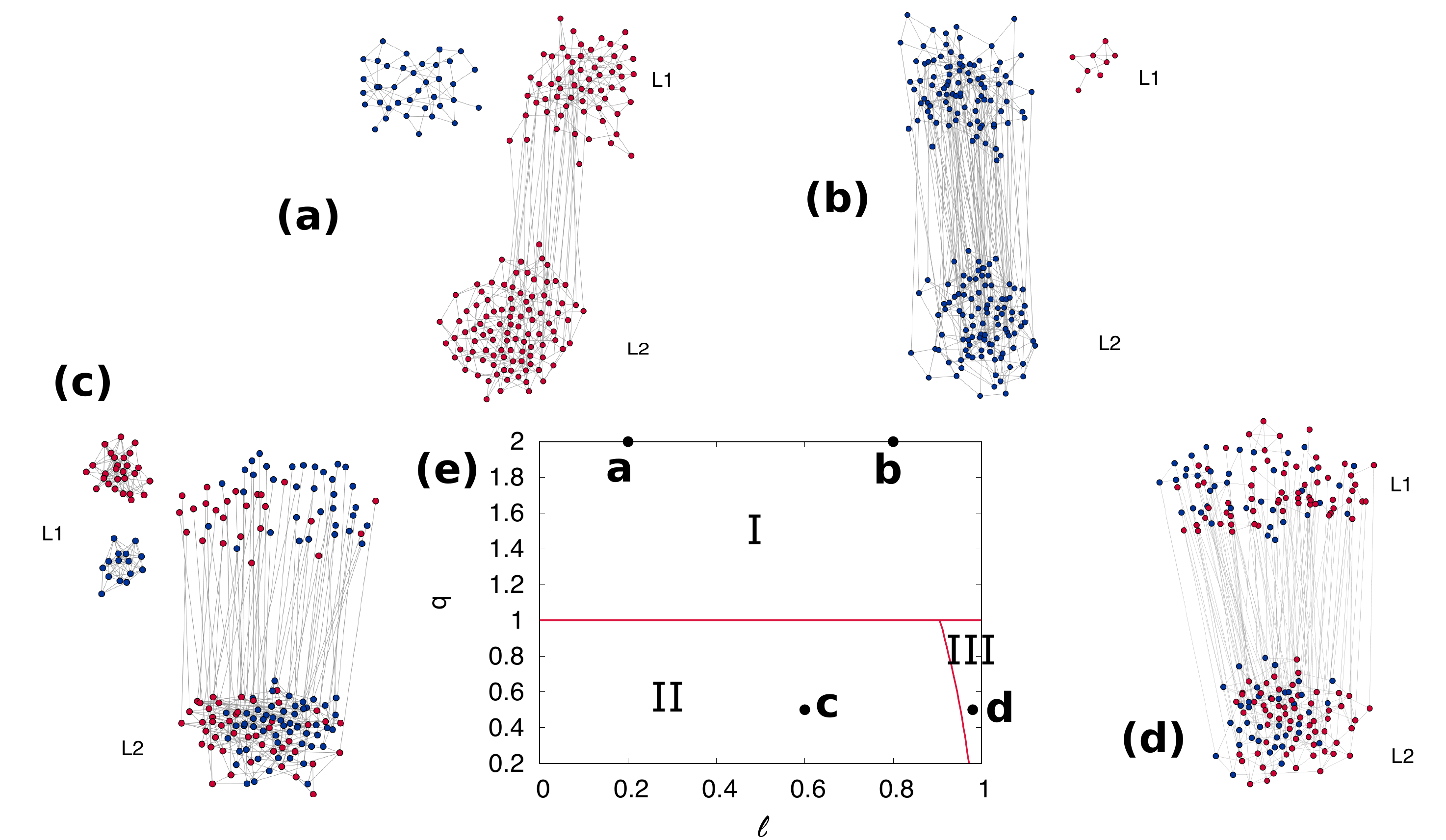}
\caption{
Phase diagram (e) for the fully asymmetric plasticity case ($p_1=1$ and $p_2=0$). 
Network configuration of (a) asymmetric fragmented phase $(q,\ell)=(2,0.2)$,
and (b) $(2,0.8)$, (c) shattered phase $(0.5,0.6)$, and (d) coexistence phase $(0.5,0.95)$.
The phase diagram is obtained numerically with multilayer initially degree regular
networks with $\langle k \rangle =4$ and $N=10^3$.
}
\label{fig:phase}
\end{figure}

For the other type of nonlinearity $q>1$, for example when $q=2$, the shattered 
phase disappears ($nc \approx 0$) and $S_1$ gradually
increases as $\ell$ increases [Fig.~\ref{fig:fully}(c,f)]. Instead of the
shattered phase, we find an asymmetric fragmented phase in which $S_1 \approx 1-S_2$
for all $\ell$ except $\ell=1$.
When $\ell=1$, we recover a consensus phase with $S_1=1$ and $S_2=0$.
The magnetization $|m|$ also increases with increasing $\ell$ since the difference
between $S_1$ and $S_2$ increases linearly. This phase with separated and asymmetric size
of two extensive clusters is also not observed in a coevolution dynamics of the 
nonlinear voter model in a single layer.

A phase diagram with respect to $\ell$ and $q$ is shown in Fig.~\ref{fig:phase}(e). We find three
different phases already described above: (I) asymmetric fragmented phase, (II) active shattered 
phase, and (III) coexistence phase. Examples of the multilayer network configuration for the 
different phases are also shown [Figs.~\ref{fig:phase}(a-d)].
When $q<1$, we find a transition at $\ell_c$ between the active shattered
and coexistence phases in the dynamic layer $L1$ while the voter layer $L2$ remains in
a dynamically active coexistence phase [Fig.~\ref{fig:phase}(c)]. When $\ell \approx 1$, the dynamic layer $L1$
also maintains a large active coexistence component due to the high degree of multiplexity [Fig.~\ref{fig:phase}(d)].
When $q>1$, the dynamic layer $L1$ exhibits two large connected clusters but with asymmetric sizes.
In addition, the size difference of the two clusters decreases linearly with $\ell$ [Figs.~\ref{fig:phase}(a,b)].
Phases (I) and (II) are not found in coevolution dynamics either in a single component network \cite{cnvm}
or in a multilayer with linear interactions \cite{marina}. Phase (III) is the analog of the dynamically 
active coexistence phase found in the single layer case, now with the same phase in the two layers. 
The difference is that in the present multilayer case this phase exists for $\ell>\ell_c$ as a 
consequence of large plasticity asymmetry, while in the single layer case it only exists below 
the fragmentation transition ($p<p_c$) as in the coexistence phase (b) in (Fig.~\ref{fig:phase_p})

\begin{figure}
\includegraphics[width=\linewidth]{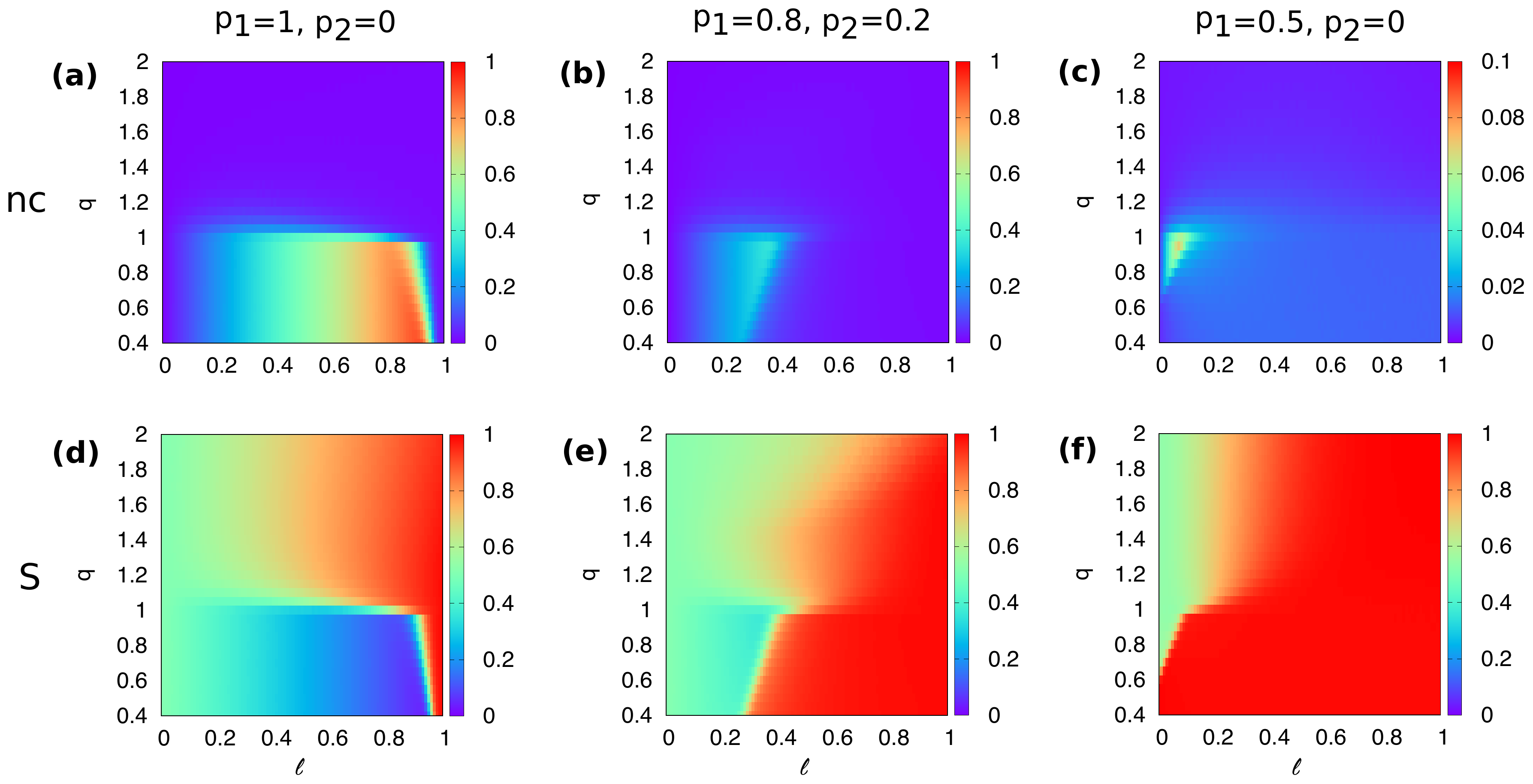}
\caption{
The relative number of components $nc$ and size $S$ of the largest component  
as a function of nonlinearity $q$ and the degree of multiplexing $\ell$ for 
partially asymmetric plasticities (a,d) $(p_1,p_2)=(1,0)$ (b,e) $(0.8,0.2)$, 
and (c,f) $(0.5,0)$ on multilayer networks with $N=10^3$..
}
\label{fig:part}
\end{figure}

The relative number of components $nc$ and the size of the largest component $S$ of the dynamic 
layer $L1$ as a function of $q$ and $\ell$ is shown in Fig.~\ref{fig:part} for the fully 
asymmetric case $p_1=1$ and $p_2=0$ and compared with results for
partially asymmetric coupling ($p_1 \ne p_2$). For the partially asymmetric cases, i.e. $(p_1,p_2)=(0.8,0.2)$
[Fig.~\ref{fig:part}(b)] and $(0.5,0)$ [Fig.~\ref{fig:part}(c)], we find that the shattered phase 
where $nc$ is nonzero is still present but in a smaller range of parameters than for the fully asymmetric case.
This finding implies that asymmetric plasticity is the source of the shattered
phase so that the area in the ($q,\ell$) parameter space where shattering occurs is maximized at fully 
asymmetric coupling. In addition, a sharp transition at $q=1$ indicates that the type of nonlinearity 
essentially determines the form of the fragmentation transition. For $q>1$, the values obtained for 
$S$ indicate that the asymmetric fragmentation phase also exists for general asymmetric values of 
the plasticity [Figs.~\ref{fig:part}(d-f)]. The range of parameters in which this phase exists 
is maximized at the fully asymmetric coupling, while for small asymmetry in the plasticity 
values a consensus phase also exists. In summary, the new phases found for the fully asymmetric 
case continue to exist when the two layers have different plasticities.

\begin{figure}
\includegraphics[width=\linewidth]{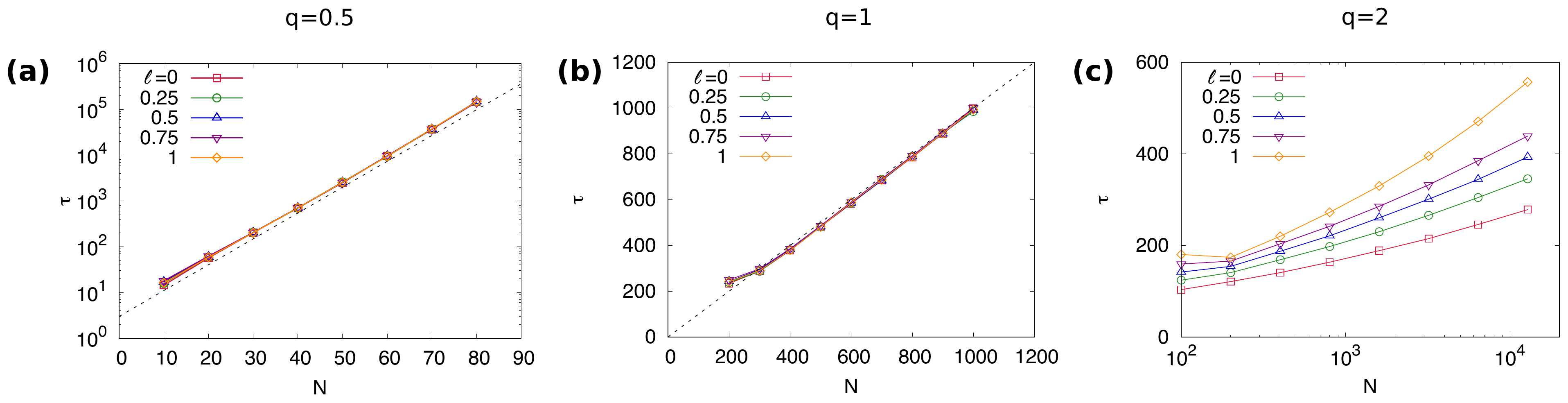}
\caption{
The characteristics time $\tau$ to reach steady state as a function of system size $N$
for (a) $q=0.5$, (b) $q=1$, and (c) $q=2$ for different $\ell$ and ($p_1,p_2$)=(1,0).
}
\label{fig:tau}
\end{figure}

Finally, we calculate the characteristic time $\tau$ to reach an absorbing state for different 
values of the nonlinear parameter $q$~(Fig.~\ref{fig:tau}). For $q=0.5$, $\tau$ increases 
exponentially with the system size $N$,[Fig.~\ref{fig:tau}(a)], so that in the thermodynamic 
limit~($N\rightarrow \infty$) we have a dynamically active coexistence phase and no absorbing 
state is reached. For finite systems, a finite size fluctuation will eventually take the system 
to an absorbing state, but due to the exponential dependence of the characteristic time on $N$, 
this is very rarely seen in our simulations and we observe dynamically active configurations 
which are extremely long lived. When $q=1$, the characteristic time grows linearly with the 
system size $N$, in the same way as in the usual voter model~[Fig.~\ref{fig:tau}(b)].
In contrast, $\tau$ increases logarithmically with $N$ for $q=2$ [Fig.~\ref{fig:tau}(c)], 
so that the absorbing state is reached in a relatively short time. The different scaling 
with $N$ of these characteristic times, for different values of the nonlinear parameter $q$, 
is consistent with previous results for a coevolving nonlinear voter model on a single 
layer network~\cite{cnvm} and also with local rewiring \cite{triadic}.

\section{Asymmetric nonlinearity in multilayer networks}

\begin{figure}
\includegraphics[width=\linewidth]{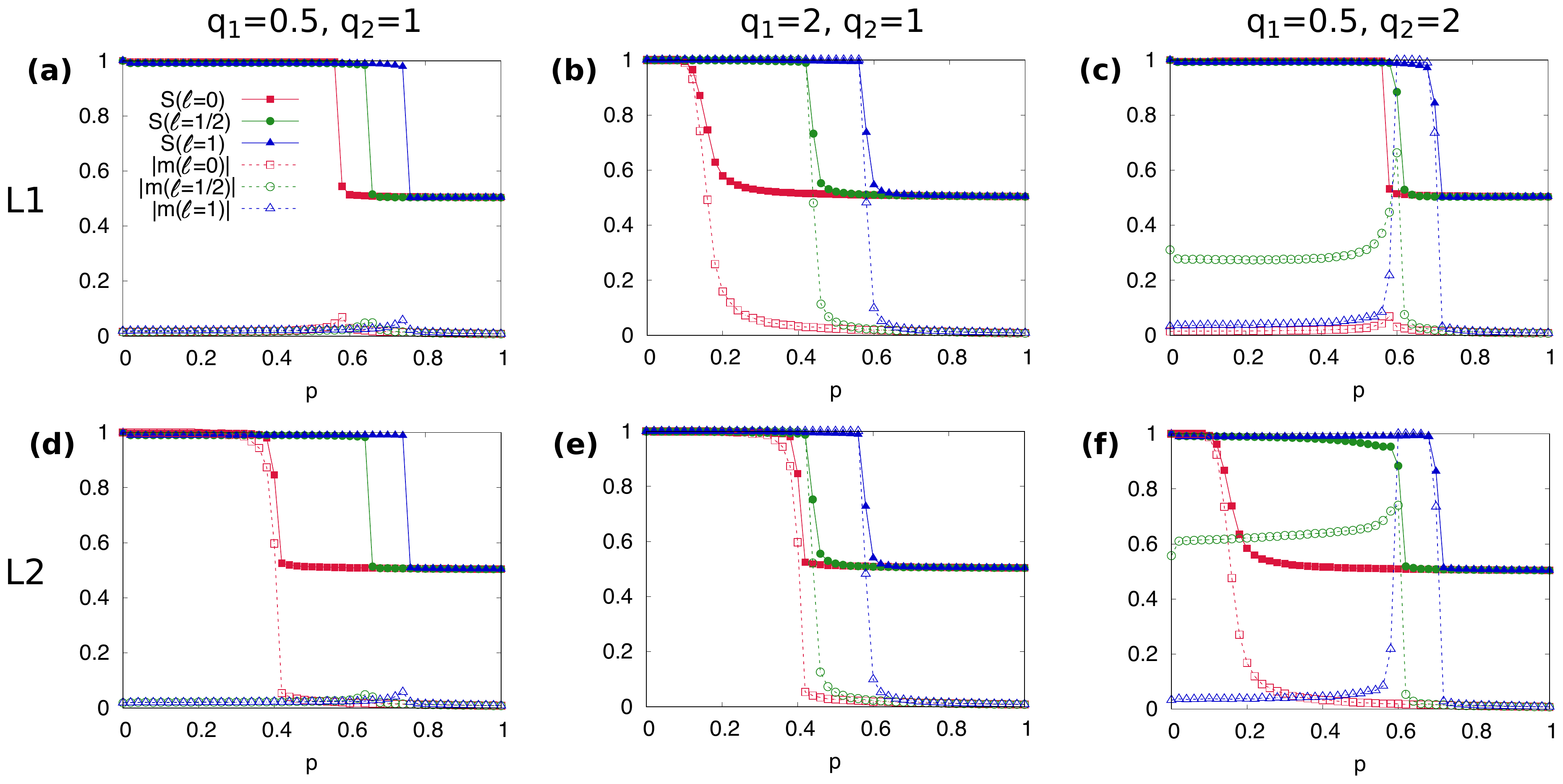}
\caption{
The size $S$ of the largest component and the absolute value $|m|$ of magnetization
of (a,d) $(q_1,q_2)=(0.5,1)$, (b,e) $(2,1)$, and (c,f) $(0.5,2)$ for different $\ell=0,0.5,1$ 
on multilayer degree regular networks with $\langle k \rangle =4$ and $N=10^4$ for each layer.
}
\label{fig:q}
\end{figure}

In this section, we consider the situation in which the two layers have a different nonlinear 
parameter $q$. Specifically, we consider three different cases $(q_1,q_2)=(0.5,1)$, $(2,1)$, 
$(0.5,2)$ with the same plasticity parameter for both layers $p=p_1=p_2$~(Fig.~\ref{fig:q}). 
For the cases $(q_1,q_2)=(0.5,1)$ and $(q_1,q_2)=(2,1)$, we find that the transition point $p_c$
is shifted for both layers with increasing $\ell$, in a similar way than we found for the 
symmetric nonlinearity case (See Fig.~\ref{fig:sym}). In this case of asymmetric nonlinearity
the layer that has slower dynamics (longer characteristics time $\tau$)
determines the steady state of the coevolving dynamics. For instance, when two layers
with $q=0.5$ and $q=1$ are coupled, the layer with $q=0.5$, with $\tau$ that grows exponentially with $N$,
dominates the dynamics, and hence the system shows a fragmentation transition between
a coexistence phase and a fragmented phase similarly to what happens when $q_1=q_2=0.5$.
In other words, in the long time limit, coevolution dynamics is shaped
by the layer taking longer to reach its final state.

We also find an anomalous fragmentation transition when two layers with different $q$
are coupled. When $(q_1,q_2)=(0.5,2)$ and $\ell=1$, there exist
two subsequent transitions: one is the transition between a coexistence phase and
a consensus phase and the other is between a consensus phase and a fragmented
phase as shown in Fig.~\ref{fig:q}(c). For intermediate $\ell=0.5$, the system 
exhibits an asymmetric active phase, that is active but $|m|\ne0$. These results 
exemplify the rich variety in phase transitions that occur in multilayer structures
with heterogeneous layer nonlinearities.

\section{Discussion}

We have studied a coevolving voter model on bilayer networks, focusing on the combined 
effect of nonlinear interactions, network plasticity and the degree of multiplexity. 
We observe a rich phase diagram with a number of new different phases and transitions.
When the two layers have the same network plasticity and nonlinear parameter, we obtain 
a fragmentation transition similar to the one obtained in a single layer~\cite{cnvm}, 
but the transition is systematically shifted to larger values of the plasticity when 
increasing the degree of multiplexity. Therefore, multiplexing prevents fragmentation~\cite{marina} 
also for a nonlinear voter model.  When the two layers have different plasticities $p$ 
but the same nonlinear parameter, we find new phases that do not exist in a coevolving 
nonlinear voter model in a single layer, namely an asymmetric fragmented phase and a 
dynamically active shattered phase. These phases are also not found in the multilayer 
version of the ordinary linear voter model. Finally, when coupling a nonlinear layer
with a linear one ($q=1$) we find that the layer with smaller nonlinearity, which is 
the one that would reach the final state in a longer time for $q_1=q_2$, dominates the 
dynamics. In addition, when coupling layers with different types of nonlinearity 
$q_1<1$ and $q_2>1$ we observe an asymmetric active phase and also, for complete 
multiplexing $\ell=1$, we observe two subsequent transitions when increasing the 
plasticity parameter: from a coexistence phase to a consensus phase, and from consensus 
to an absorbing fragmented phase.

\ack
We acknowledge financial support from Agencia Estatal de Investigacion (AEI, Spain) and
Fondo Europeo de Desarrollo Regional under Project ESoTECoS Grant
No. FIS2015-63628-C2-2-R (AEI/FEDER,UE) and from Agencia Estatal de Investigacion (AEI, Spain)
through the Maria de Maeztu Program for Units of Excellence in R\&D (MDM-2017-0711).
This work was supported by the National Research Foundation of Korea (NRF) grant
funded by the Korea government (MSIT) (No. 2018R1C1B5044202).


\section*{References}
\begin{thebibliography}{99}
\bibitem{vidal} Vidal M, Cusick M E, Barab\'asi A L 2011
	{\it Cell} {\bf 144} 986
\bibitem{white} White J G, Southgat E, Thomson J N, Brenner S 1986
	{\it Phil. Trans. Royal Soc. London B} {\bf 314} 1
\bibitem{borge} Borge-Holthoefer
	{\it et al} 2011 {\it PLoS ONE} {\bf 6(8)} e23883
\bibitem{verbrugge} Verbrugge 1979
	{\it Social Forces} {\bf 57} 1286
\bibitem{domenico} De Domenico M, Sol\'e-Ribalta A, G\'omez S, Arenas A 2014
	{\it Proc. Natl. Acad. Sci. USA} {\bf 111} 8351
\bibitem{little} Little R G 2002
	{\it J. Urban Technology} {\bf 9} 109
\bibitem{rosato} Rosato V, Issacharoff L, Tiriticco F, Meloni S, Porcellinis S, Setola R 2008
	{\it J. Critical Infrast.} {\bf 4} 63
\bibitem{robustness} Min B, Yi S-D, Lee K-M, Goh K-I 2014
	{\it Phys. Rev. E} {\bf 89} 042811

\bibitem{kivela} Kivel\"a M, Arenas A, Barthelemy M, Gleeson J P, Moreno Y, Porter M A 2014
	{\it J. Complex Netw.} {\bf 2} 203
\bibitem{leicht} Leicht E A and D'Souza R M 2009
	arXiv:0907.0894
\bibitem{buldyrev} Buldyrev S V, Parshani R, Paul G, Stanley H E and Havlin S 2010
	{\it Nature} {\bf 464} 1025
\bibitem{lee} Lee K-M, Min B, Goh K-I 2015
	{\it Eur. Phys. J. B} {\bf 88} 48
\bibitem{boccaletti} Boccaletti S, Bianconi G, Criado R, del Genio C I, G\'omez-Garde\~nes, Romance M,
	Sendi\~na-Nadal I, Wang Z, Zanin M 2014 {\it Phys. Rep.} {\bf 544} 1
\bibitem{wang} Wang Z, Wang L, Szolnoki A, Perc M 2015
	{\it Eur. Phys. J. B} {\bf 88} 124
\bibitem{marina2} Diakonova M, Nicosia V, Latora V, San Miguel M 2016
	{\it New J. Phys.} {\bf 18} 023010

\bibitem{son} Son S-W, Grassberger P, Paczuski M 2011
	{\it Phys. Rev. Lett.} {\bf 107} 195702
\bibitem{viability} Min B and Goh K-I 2014
	{\it Phys. Rev. E} {\bf 89} 040802
\bibitem{moreno} G\'omez S, D\'iaz-Guilera A, G\'omez-Garde\~nes J, P\'erez-Vicente C J, Moreno Y, Arenas A 2013
	{\it Phys. Rev. Lett.} {\bf 81} 056105
\bibitem{switching} Min B, Gwak S-H, Lee N, Goh K-I 2016
	{\it Sci. Rep.} {\bf 6} 21392
\bibitem{zheng} Min B and M. Zheng 2018
	{\it PLoS One} {\bf 13(4)} e0195539
\bibitem{rescue} Vazquez F, Serrano M A, San Miguel M 2016
	{\it Sci. Rep.} {\bf 6} 29342
\bibitem{czaplicka} Czaplicka A, Toral R, San Miguel M 2016
	{\it Phys. Rev. E} {\bf 94} 062301
\bibitem{brummitt} Brummitt C D, Lee K-M, Goh K-I 2012
	{\it Phys. Rev. E} {\bf 85} 045102


\bibitem{marina} Diakonova M, San Miguel M, Egu\'iluz V M 2014
	{\it Phys. Rev. E} {\bf 89} 062818
\bibitem{amato} Amato R, Kouvaris N, San Miguel M, Diaz-Guilera A 2017
	{\it New J. Phys.} {\bf 19} 123019
\bibitem{artime} Artime O, Fernández-Gracia J, Ramasco José J, San Miguel M 2017
	{\it Sci. Rep.} {\bf 7}, 7166



\bibitem{klimek} Klimek P, Diakonova M, Eguiluz V. M., San Miguel M, Thurner S 2016
	{\it New J. Phys.} {\bf 18} 083045
\bibitem{lugo} Lugo Haydee and San Miguel M 2015
	{\it Sci. Rep.} {\bf 5} 7776
\bibitem{szolnoki} Szolnoki A and Perc M 2013
	{\it New J. Phys.} {\bf 15} 053010
\bibitem{wang2} Wang Z, Wang L and Perc M 2014
	{\it Phys. Rev. E} {\bf 89} 052813
\bibitem{battiston} Battiston F, Nicosia V, Latora V, San Miguel M 2017
	{\it Sci. Rep.} {\bf 7} 1809
\bibitem{perc} Perc M and Szolnoki A 2010
	{\it BioSystems} {\bf 99} 109-125

\bibitem{zimmerman} Zimmerman M G, Egu\'iluz V M, San Miguel M 2004
	{\it Phys. Rev. E} {\bf 69} 065102
\bibitem{vazquez} Vazquez F, Egu\'iluz V M, San Miguel M 2008
	{\it Phys. Rev. Lett.} {\bf 100} 108702
\bibitem{holme} Holme P and Newman M E J 2006
	{\it Phys. Rev. E} {\bf 74} 056108

\bibitem{cnvm} Min B and San Miguel M 2017
	{\it Sci. Rep.} {\bf 7} 12864
\bibitem{triadic} Raducha T, Min B, San Miguel M 2018
	{\it Europhys. Lett.} {\bf 124} 3001
\bibitem{castellano} Castellano C, Mu\~noz, M A, Pastor-Satorras R 2009
	{\it Phys. Rev. E} {\bf 80} 041129
\bibitem{nyczka} Nyczka P, Sznajd-Weron, K and Cislo J  2012
	{\it Phys. Rev. E} {\bf 86} 011105
\bibitem{mf} Jedrzejewski A 2017
	{\it Phys. Rev. E} {\bf 95} 012307
\bibitem{peralta} Peralta A F, Carro A, San Miguel M, Toral R 2018
	{\it Chaos} {\bf 28} 075516

\bibitem{nowak} Nowak A, Szamrej J, Latan\'e B 1990
	{\it Psychological Review} {\bf 97} 362
\bibitem{nettle} Nettle D 1999
	{\it Lingua} {\bf 108} 95
\bibitem{abrams} Abrams D M and Strogatz S H 2003
	{\it Nature} {\bf 424} 900
\bibitem{jstat} Vazquez F, Castello X, San Miguel M 2010
 	{\it J. Stat. Mech.} {\bf 04} P04007
\bibitem{castello} Castello X, Loureiro-Porto, L and San Miguel M 2013
 	{\it International Journal of the Sociology of Language} {\bf 221} 21



\end{thebibliography}
\end{document}